\documentclass[traditabstract]{aa} 
\usepackage{graphicx}
\usepackage{txfonts}
\usepackage{supertabular}
\usepackage{textcomp}
\usepackage{booktabs}
\usepackage{lscape}
\usepackage{xspace}
\usepackage{tabularx}
\usepackage{txfonts}
\begin{document}
\title{
  Photometric AGN reverberation mapping -- an efficient
  tool for 
  BLR sizes, black hole masses and host-subtracted AGN luminosities
}

	\titlerunning{Photometric AGN reverberation mapping}

	\author{
		M. Haas
		\inst{1}
		\and
		R. Chini
		\inst{1,2}
		\and
                M. Ramolla
		\inst{1}
		\and
                F. Pozo Nu\~nez
		\inst{2}
		\and
                C. Westhues
		\inst{1}
		\and
                R. Watermann
		\inst{1}
		\and
                V. Hoffmeister
		\inst{1}
		\and
                M. Murphy
		\inst{2}
		}

	\institute{Astronomisches Institut, Ruhr--Universit\"at Bochum,
		Universit\"atsstra{\ss}e 150
                , 44801 Bochum, Germany
		\and
                Instituto de Astronomia, Universidad Cat\'{o}lica del
                Norte, Avenida Angamos 0610, Casilla
                1280 Antofagasta, Chile\\}
	\authorrunning{M. Haas et al.}

	\date{Received ; accepted }


	\abstract{
          Photometric reverberation mapping
          employs a wide bandpass to measure the AGN continuum
          variations and a
          suitable band, usually a narrow band (NB),
          to trace the echo of an emission
          line in the broad 
          line region (BLR). 
          The narrow band catches both the 
          emission line and the underlying continuum,
          and one needs to extract the pure emission line light curve.
          We performed a test on two local AGNs, PG0003+199 (=Mrk335) 
          and Ark120, observing well-sampled 
          broad- ($B$, $V$) and narrow-band light curves 
          with the robotic 15cm telescope VYSOS-6 on Cerro Armazones, 
          Chile. 
          In PG0003+199, H$\alpha$ dominates the flux in the NB by 
          85\%, allowing us to measure the time lag of H$\alpha$ against $B$  
          without the need to correct for the continuum contribution.
          In Ark120, 
          H$\beta$ contributes only 50\% to the flux in the NB.
          The cross correlation of the $B$ and NB light curves shows 
          two distinct peaks of similar strength, 
          one at lag zero from the autocorrelated 
          continuum and one from the emission line
          at $\tau_{cent}$  =  47.5 $\pm$ 3.4  days.
          We constructed a synthetic H$\beta$ light curve, by subtracting a
          scaled $V$ light curve, which traces the continuum, 
          from the NB light curve.
          The cross correlation of this synthetic H$\beta$ light curve
          with the $B$ light curve 
          shows only one major peak at  
          $\tau_{cent}$ = 48.0 $\pm$ 3.3 days, while
          the peak from the autocorrelated continuum at lag zero is absent. 
          We conclude that, 
          as long as the emission line contributes at least 50\% 
          to the bandpass, 
          the pure emission line light curve can be 
          reconstructed from photometric monitoring
          data so that the time lag can be measured. 
          For both objects the lags we find are consistent 
          with spectroscopic reverberation results.
          While the dense sampling (median 2 days) enables us to determine 
          $\tau_{cent}$ 
          with small ($\sim$10\%) formal errors,   
          we caution that gaps in the light curves may lead to 
          much larger systematic uncertainties. 
          We calculated a virial black hole mass for both objects 
          in agreement with
          literature values, by combining $R_{\rm BLR}$ ($\tau_{cent}$)
          from photometric monitoring with the velocity dispersion 
          of a single contemporaneous spectrum. 
          We applied
          the flux variation gradient method to estimate the 
          host galaxy contribution in the apertures used and thus the 
          host-subtracted restframe 5100\AA~  luminosity $L_{AGN}$ 
          during the time of our monitoring campaign.  
          At the AGN/host contrast of our sources ($L_{AGN}/L_{host} \sim 1$)
          the uncertainty of $L_{AGN}$ is $\sim$10\%. 
          For both sources, our $L_{AGN}$ differs significantly from previous 
          estimates, placing both sources $\sim$50\% closer to the 
          $R_{\rm BLR} - L_{AGN}$ relation. 
          Our results suggest that the  scatter 
          in the current $R_{\rm BLR} - L_{AGN}$ relation is largely 
          caused by uncertainties in $R_{\rm BLR}$ due to 
          undersampled light curves and by uncertainties in the 
          host-subtracted AGN luminosities inferred so far.
          If the scatter can be 
          reduced sufficiently, then two quasar samples matching in 
          $R_{\rm BLR}$ should match also in intrinsic 
          $L_{AGN}$, independent of 
          redshift, thus offering the prospect to probe cosmological models.
          Photometric reverberation
          mapping opens the door to efficiently measure hundreds of
          BLR sizes and host-subtracted AGN luminosities even 
          with small telescopes, but also routinely with upcoming large 
          survey telescopes like the LSST. 
	}

		\maketitle
%

\section{Introduction}

Reverberation mapping (Blandford \& McKee 1982) has revolutionised our
understanding of active galactic nuclei (AGN) during the past
decades. Spectroscopic monitoring measures 
the mean time lag $\tau$ between brightness changes of
the triggering optical-ultraviolet continuum (from the compact
accretion disk around the black hole) and the emission line brightness
echo of the BLR gas clouds further out; the mean BLR size is then
$R_{BLR} = \tau \cdot c$, where $c$ is the speed of light. 
The BLR size as measured by the centroid $\tau_{cent}$ of
the cross correlation of light curves 
represents a luminosity-weighted average 
(Koratkar \& Gaskell 1991, Penston 1991).
Evidence
exist that the BLR clouds are essentially in virialised motion around
the central black hole (Peterson et al. 2004 and references
therein). Combining $R_{\rm BLR}$ with velocity dispersion
$\delta v$ of the emission line, the virial black hole mass can be
estimated as $M_{\rm BH} = f \cdot R_{\rm BLR} \cdot \delta v / G$,
where $f$ is a scaling factor in the order of unity to account for the
geometry of the BLR and $G$ is the gravitational constant. 

Reverberation based BLR sizes and black hole masses have been
determined so far for only about 45 low- and intermediate-luminous AGN
(Wandel et al. 1999, Kaspi et al. 2000, Peterson et al. 2004, Bentz et
al. 2010). Six years of reverberation mapping of six 
luminous high-redshift ($z > 2$) quasars, using the 9.2\,m
Hobby-Eberly-Telescope in Texas, resulted in a tentative $R_{\rm BLR}$ 
determination of only one quasar, while in the other five sources the
existence of pronounced variation patterns could be established 
(Kaspi et al. 2007). 

Reverberation spectra usually require observations with at least a
2\,m-class telescope. The BLR sizes range between a few and several
hundred light days and the time spans become even longer in the
observer's frame due to the time dilation factor $1 + z$. To match
distinct echo features, light curves over a few months to several
years are needed. Consequently, spectroscopic reverberation mapping is
resource expensive and it becomes prohibitive at high
redshift. Therefore,  more efficient observing strategies
are desired. 

Another issue with respect to the BLR size -- AGN luminosity relationship 
is how to determine the AGN luminosity free of host 
galaxy contributions. 
While high spatial resolution images have been invoked 
to model and subtract the host contribution from the total fluxes 
(e.g. Bentz et al. 2009), 
the flux variation gradient method (Choloniewski 1981)
provides an efficient approach
which we will inspect here as well. 

We here revisit photometric reverberation mapping. 
It employs wide pass bands to
trace the AGN continuum and suitable narrow-bands to trace the echo of
the BLR lines. The advantage is that photometric monitoring can be 
obtained much faster and with small telescopes. Furthermore, while
long-slit spectroscopic monitoring allows for simultaneous measurement
of typically one or two calibration stars placed accurately on the
spectrograph slit, one may expect that the large number of
non-variable calibration stars on the same images and of similar
brightness as the AGN facilitates to attain photometric monitoring
with high precision. 

About 40 years ago, narrow-band AGN monitoring of emission lines has
been performed, for instance for NGC\,3516 and NGC\,4151
(Cherepashchuk \& Lyutyi 1973). 
Despite the coarse ($>$10 days) time sampling,
these observations discovered a lag between H$\alpha$ and
the continuum of 25 and 30 days for NGC\,3516 and NGC\,4151,
respectively. 
However, spectroscopic monitoring with better time sampling
finds a lag of $13 \pm 4$ days for NGC\,3516 (Peterson et al. 2004) 
and  $6.6 \pm 1.1$ days for NGC\,4151 (Bentz et al. 2006). 
The discrepancy between the photometrically and 
spectroscopically determined lags is most likely due to insufficient 
time sampling of the early photometric monitoring.  

To explore the capability and achievable
accuracy of photometric reverberation mapping, 
we selected two local AGNs, the relatively clear-cut case
PG0003+199 (Mrk\,335) at redshift $z = 0.0258$ and the more
challenging case Ark\,120 at $z = 0.0327$. Spectroscopic reverberation
results are available for comparison (Peterson et al. 1998a, 2004).

\section{Data}

We performed a monitoring campaign between August 2009 and March 2010
using the robotic 15\,cm telescope VYSOS-6 of the Universit\"atssternwarte
Bochum on Cerro Armazones, the future location of the ESO Extremely
Large Telescope in Chile. VYSOS-6 is equipped with a $4096 \times
4096$ pixel CCD yielding a field-of-view of $2.7^{\circ}$, and seven
broad- and narrow-band filters.\footnote{Since August 2010, 
VYSOS-6 consists of two 15\,cm telescopes on the same mount, 
equipped with 14 broad- and narrow-band filters.} 

\begin{figure}
  \centering
  \includegraphics[width=\columnwidth]{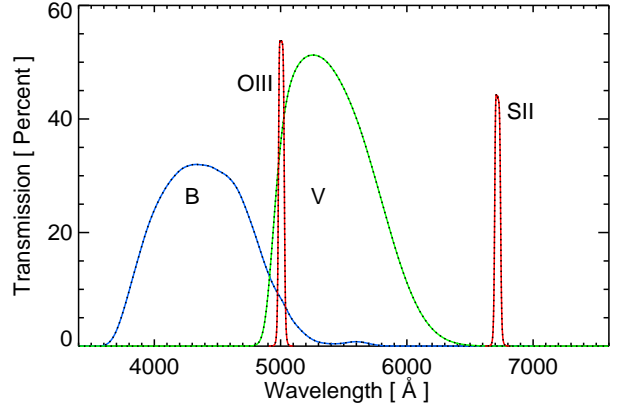}
  \caption{ Effective transmission of the filters convolved with  
    the quantum efficiency of the ALTA U16M CCD camera. 
  }
  \label{fig_transmission}
\end{figure}

We obtained light curves with a median sampling of 2 days in the
$B$-band (Johnson, $4330 \pm 500$\,\AA), the redshifted H$\alpha$ (NB
$6721 \pm 30$\,\AA \,at $z = 0.0258$) and H$\beta$ lines (NB $5007 \pm
30$\,\AA \,at $z = 0.0327$). For Ark\,120 we also obtained a light
curve in the $V$-band (Johnson, $5500 \pm 500$\,\AA). 
In addition, we observed PG0003+199 in both $B$ and $V$ 
on July 25$^{th}$ 2010 and June 26$^{th}$ 2011.
Figure~\ref{fig_transmission} shows the effective 
transmission of the filters used here.
To perform absolute photometric calibration, each night 
we observed standard stars
in the fields SA092, SA095, SA111 from Landoldt (2009). 
For both targets, a single contemporaneous spectrum was
obtained with CAFOS at the 2.2\,m telescope on Calar Alto, Spain, 
with a slit width of 1$\farcs$54. 

We reduced all data in a standard manner, using IRAF and custom
written tools. 
Because the flux calibration using the standard star fields 
introduces additional errors in the light curves, 
we created relative light curves (in normalised flux units)
using 20-30 non-variable stars located on the same images within
30$\arcmin$ around the AGN and of similar brightness as the AGN. 
For the analysis of time lags we used the mean and standard deviation 
of these relative light curves. For the photometric analysis 
(to obtain the AGN luminosities), we kept the shape of 
the mean light curves fixed and calibrated them by a least-squares fit
to the photometry derived from the standard star fields. 
Source parameters and photometry results are summarized in 
Table~\ref{table1}. 

\begin{figure}
  \centering
  \includegraphics[angle=90,width=\columnwidth]{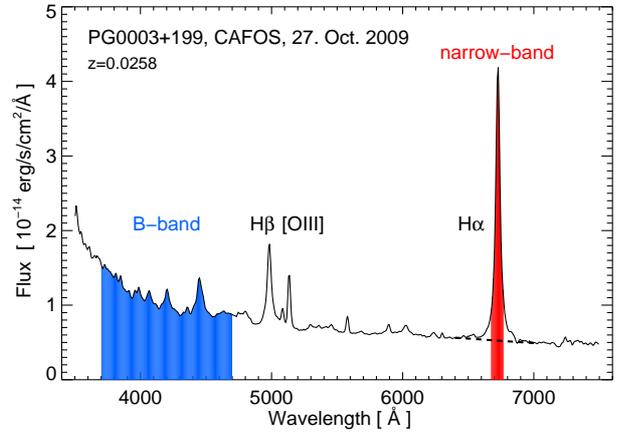}
  \caption{
    Observed spectrum of PG0003+199. The $B$-band (blue shaded) traces
    essentially the
    AGN continuum, while the narrow-band (red shaded) is dominated by
    the H$\alpha$ line. 
      Note that the blue and red shaded areas are a schematic 
      illustration and
      that quantitative calculations were made using 
      the  transmission curves shown in 
      Figure~\ref{fig_transmission}.
  }
  \label{fig_spectrum_pg0003}
\end{figure}

\section{Results}

\subsection{Results on PG0003+199}

PG0003+199 is essentially a point-like
low-luminosity quasar (Narrow Line Seyfert-1) without a bright
extended host. The H$\alpha$ line is strong and well covered by the 
narrow-band filter. This makes PG0003+199 a clear-cut test case. 

\subsubsection{Spectrum}

Fig.~\ref{fig_spectrum_pg0003} shows the spectrum of PG0003+199. The
contribution of higher order Balmer lines (and that of the host
galaxy) to the $B$-band is negligible, thus the $B$-band is dominated
by the AGN continuum. Taking into account the spectral resolution, the
observed H$\alpha$ line dispersion $\sigma = 1300$\,km/s reduces to an
intrinsic value $\sigma = 870$\,km/s. The narrow-band filter
effectively covers the line between velocities  $-2800$\,km/s and
$+1800$\,km/s, so that at least 95\% of the line flux is contained in the
band pass, as we determined after line profile deconvolution. 

The contribution of both the [N II]\,6583\,\AA \,and narrow-line
H$\alpha$ flux is predicted to be smaller than 40\% of the [O III]
5007\,\AA~ emission (Bennert et al. 2006), hence negligible
($<10\%$). The continuum underneath the emission line is small,
contributing to only 15\% of the total flux in the band
pass. Therefore, the narrow-band light curves will be dominated by the
H$\alpha$ echo of the BLR gas clouds. Comparison with simulated line
profiles of echo models (Welsh \& Horne 1991, Horne et al. 2004)
ascertains that the 5\% line flux outside the band pass has only a
marginal effect ($< 2$\%) on the BLR size determination: The
narrow-band echo may miss only a small fraction ($< 20$\%) of the
innermost part of the BLR, namely that part which exhibits the fastest
line-of-sight velocity, while the innermost gas clouds moving closer
along the sky plane exhibit a modest line-of-sight velocity and are
therefore contained in the narrow-band. 

\begin{figure}
  \centering
  \includegraphics[angle=90,width=\columnwidth]{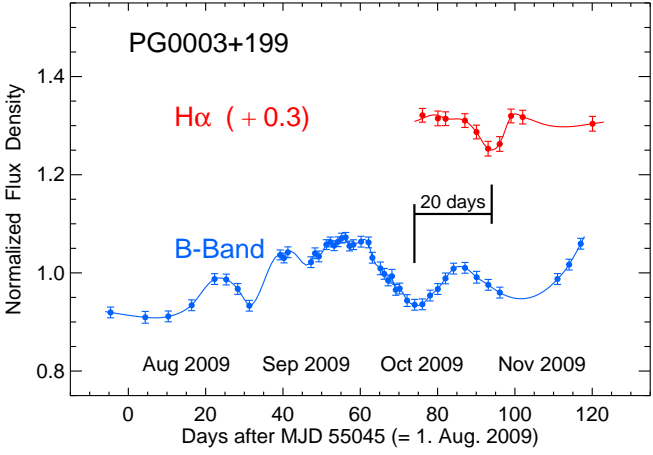}
  \caption{
      $B$-band and H$\alpha$  light curves of PG0003+199. The
      H$\alpha$ observations were started in October 2009 after
      recognizing the pronounced $B$-band variations. 
  }
  \label{fig_lightcurve_pg0003}
\end{figure}

\subsubsection{Light curves and BLR size}
\label{section_pg0003_lc}

Fig.~\ref{fig_lightcurve_pg0003} depicts the light curves of
PG0003+199. The AGN continuum as traced by the $B$-band increases 
gradually from
August 2009 to a peak around end of September, followed by a steep
trough by 20\% at MJD $\sim$55118 (= 55045 + 73). Also the H$\alpha$
light curve has a steep $\sim$10\% trough with a delay of about 20
days. 

\begin{figure}
  \centering
  \includegraphics[angle=90,width=\columnwidth,clip=true]{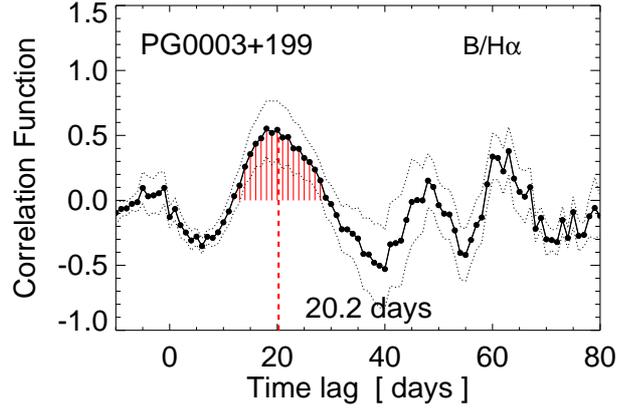}
  \caption{ 
    Cross correlation of the $B$ and H$\alpha$ light curves of
    PG0003+199, taking all data points. 
      The dotted lines indicate the error range  ($\pm 1\sigma$) around the 
      cross correlation.
    The red shaded area markes the range to calculate the 
    centroid $\tau_{cent}$ of the major
    correlation feature, yielding a lag of 20.2 days (vertical dashed line).
  }
  \label{fig_dcf_pg0003}
\end{figure}

We used the discrete correlation function (DCF, Edelson \& 
Krolik 1988) to cross correlate the H$\alpha$ and $B$-band light
curves. 
The cross correlation shows a major peak with a lag of 20.2 days as
defined by the centroid $\tau_{cent}$ (Fig. \ref{fig_dcf_pg0003}).
Two smaller correlation peaks around lag 50 days and 65
days are present; they are caused by the $B$-band troughs at
begin and mid of September 2009 (Fig.~\ref{fig_lightcurve_pg0003}). 
We here do not consider these two lags further, 
because spectroscopic reverberation predicts a lag of less 
than 30 days (Peterson et al. 2004). 

\begin{figure}
  \centering
  \includegraphics[angle=90,width=\columnwidth,clip=true]{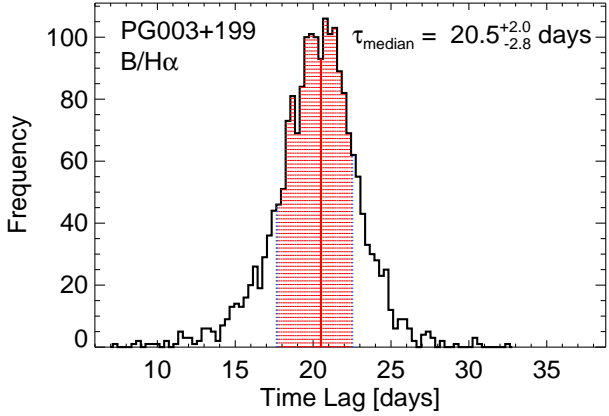}
  \caption{
    Result of the FR/RSS error analysis of the 
    cross correlation of the $B$ and H$\alpha$ light curves of PG0003+199. 
    The histogram shows the distribution of
    $\tau_{cent}$ for 2000 flux randomised and randomly selected
    subset light curves. 
    The median of this distribution is taken as 
    the final value of the lag $\tau_{cent}$.   
    The red shaded area marks the 68\%  confidence range used to calculate the
    errors of $\tau_{\rm cent}$. 
  }
  \label{fig_hist_pg0003}
\end{figure}

To determine the lag uncertainty, we applied the flux randomization 
and random subset selection method (FR/RSS, Peterson et al. 1998b).
From the observed light curves we created 2000 randomly selected
subset light curves, each containing 63\% of the original data points,
and randomly altering the flux value of each data point consistent 
with its (normal-distributed) measurement error.
Then we cross correlated the 2000 pairs of subset light curves and
computed the centroid $\tau_{cent}$. 
Fig. \ref{fig_hist_pg0003} shows the
distribution of the 2000 $\tau_{cent}$ values which yields the
median lag $\tau_{cent}$ and the 68\% confidence range. 
From this distribution we obtain 
a lag $\tau_{cent}$ = 20.5 $^{+2.0}_{-2.8}$ days. 
Thus, the mean of positive and negative uncertainty
of $\tau_{cent}$ is about 12\% (2.4/20.5).\footnote{The observed
  H$\alpha$ light curve has only 10 data points, three of which are 
in the trough. While a subset of exactly  
63\% data points would require to take 6.3 data points, 
we selected 7 random data points. 
This $\sim$10\% larger number of data points may bias the errors to
somewhat smaller values. On the other hand, we find the largest errors
for those randomly selected subset light curves, 
where all three data points in the trough are omitted. 
In this case the 
information about the existence of a trough is completely lost. 
To counterbalance this effect, we selected 7 
instead of 6.3 data points.
} 
Correcting for the time dilation (factor $1+z$ at $z=0.0258$) 
yields a rest frame lag of 20.0 $^{+2.0}_{-2.7}$  days.

We did not find H$\alpha$ lags of PG0003+199 in the literature,
but we can compare our result with that of the H$\beta$ line for which
spectroscopic reverberation mapping found a rest-frame lag of $15.7 \pm 3.7$
days averaged over several epochs 
(Peterson et al. 2004, Bentz et al. 2009).
This H$\beta$ lag is $\sim$30\% smaller than our H$\alpha$ lag of 20.0 days.
One possible explanation could be that the 
BLR size is smaller for higher excitation emission lines.
Kaspi et al. (2000) found in 
the spectroscopic reverberation data of 17 PG quasars
 $R_{{\rm H}\alpha} = s \cdot R_{{\rm H}\beta}$ 
with a scaling factor $s = 1.19$, but a high uncertainty of $\pm 0.23$. 
On the other hand, Kollatschny (2003a) 
provides clear evidence for the BLR stratification  of
Mrk\,110 ($s = 1.37$, his Table\,1). 
Recently, the Lick AGN
Monitoring Program of 11 low-luminosity AGN found 
$s = 1.54 \pm 0.4$ (Bentz et al. 2010). 
It could be that the BLR stratification and the scaling factor $s$  
depends on the luminosity.
The luminosity of PG0003+199 is similar to that of Mrk\,110.
If the  value $s=1.37$ of Mrk\,110 
holds for PG0003+199, then the H$\beta$
lag of  PG0003+199 translates to an H$\alpha$ lag of 
15.7 $\times$ 1.37 = 21.5 days, hence agrees 
within 8\%  with the restframe H$\alpha$ lag of 20.0 days 
from our measurement. 

\subsubsection{Black hole mass}

Next we try to determine the mass of the central black hole.
While spectroscopic reverberation data enable one to use the velocity
information from the rms spectra, i.e. from the variable portion of
the emission lines, several attempts have been performed 
to determine the mass of the central black hole using single epoch
spectra (e.g. Vestergaard 2002, Woo et al. 2007, Denney et al. 2009a).
Thereby the reported uncertainties of the velocity dispersion range between
15\% and 25\%. 
Combining for PG0003+199 the intrinsic line dispersion from our
single spectrum
($870$\,km/s, with 25\% uncertainty adopted)
and the time delay (20.0 $\pm$ 2.4  days)
yields a virial black
hole mass $M_{\rm BH}  = 2.8 \pm 1.1 \cdot 10^{6}\, M_{\odot}$, 
in agreement
with the value of $2.8 \pm 0.8 \cdot 10^{6}$\,M$_\odot$ derived via
spectroscopic reverberation mapping by Peterson et al. (2004). 

\subsubsection{Host-subtracted AGN luminosity}

To determine the AGN luminosity free of host galaxy contributions, 
we applied the flux variation gradient (FVG) method proposed by
Choloniewski (1981). In this method 
the $B$ and $V$ data points obtained in the same night 
through the same apertures 
are plotted  in a $B$-flux versus $V$-flux diagram.
Note that fluxes (e.g. in units of mJy) 
are plotted and not magitudes. 
The important feature is that the flux variations follow a linear
relation with a slope $\Gamma_{BV}$ given by the host-free AGN continuum  
(e.g. Choloniewski 1981, Winkler et al. 1992, Winkler 1997). 

In the flux-flux diagram the host galaxy -- including the contribution of line
emission from the narrow line region (NLR) -- 
 lies on the AGN slope somewhere toward its fainter end.  
This has been demonstrated for a sample of 11 nearby AGN 
(neither containing PG0003+199 nor Ark120) by Sakata et al. (2010), 
using monitoring data in combination with Hubble Space Telescope
and multi-band ground-based images from the MAGNUM telescope. 
For 8$\farcs$3 aperture, 
they derived host galaxy colours in the range 0.8 $ < B-V <$ 1.1 
(their Table 6), which agree with the typical colours of a bulge 
or an elliptical galaxy. 
This colour range corresponds to a host slope in the range 
$0.4 < \Gamma_{BV}^{host} < 0.53$. 
The contribution of the NLR
emission lines to the $B$ and $V$ bands is less than 10\% of the host
flux and has a similar color as the host galaxy 
(cf. Tables 6 and 8 of Sakata et al. 2010). 
We note that dust enshrouded nuclear starbursts, 
which often accompany the AGN phenomenon, have red colors ($B-V \sim 1$). 
They would be unresolved on HST images and could explain the small 
deviations of the host position from the AGN slope found 
in three cases by Sakata et al. (2010).
Therefore, 
by host we here denote the host galaxy including NLR AGN lines 
and starbursts. 

The host slopes  pass through the origin. 
Because the host slopes  are  flatter
than  typical AGN slopes ($\Gamma_{BV} \sim 1$),  
the intersection of the two slopes should occur in a well defined range. 
Winkler et al. (1992) showed for NGC\,3783 that 
this intersection range allows one to determine the host flux 
contribution and
to calculate a host-subtracted AGN luminosity at the time of
the monitoring campaign -- even without the need for high spatial 
resolution images. 
Next (and in Sect. \ref{section_ark120_fvg}) 
we test this procedure on our data. 
Table~\ref{table2} summarizes the 
results of the FVG diagnostics, 
which are derived as follows:

\begin{figure}
  \centering
  \includegraphics[angle=0,width=\columnwidth]{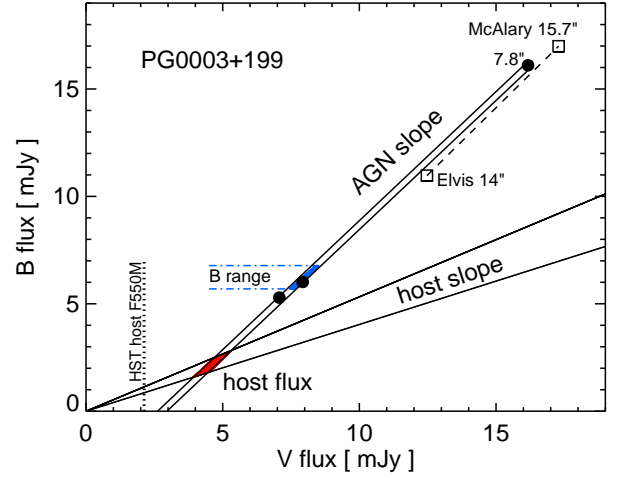}
  \caption{ 
    $B$ versus $V$ fluxes of PG0003+199, measured in a 7$\farcs$8
    aperture (black dots). 
    Errors are within the size of the symbols.
    A linear least-squares fit to the three data points yields the
    range for the AGN slope plotted by the two steep lines.
    This slope is corroborated by the slope (dashed line) 
    obtained with larger apertures 14$\arcsec$ (Elvis) 
    and 15$\farcs$7 (McAlary). 
    The range of host slopes plotted by the two shallow lines 
    is taken from Sakata et al. (2010).
    The intersection area of the AGN and host slopes (red shaded) 
    defines the host contribution. 
    The blue shaded area marks the intersection of the AGN slope with the 
    range of $B$ fluxes during our 
    Aug-Nov 2009 monitoring campaign. 
    This area allows us to infer the range of $V$ fluxes during 
    the  monitoring campaign.  
    The vertical dotted line marks the host flux found 
    in an aperture of $5\arcsec \times 7\farcs6$ 
    via modelling of HST F550M images by Bentz et al. (2009).  
    All data are corrected for galactic foreground extinction. 
  }
  \label{fig_fvg_pg0003}
\end{figure}

During our monitoring
campaign of PG0003+199  in Aug-Nov 2009, we obtained only $B$ fluxes, 
but we have three data points at hand for the $B$ versus $V$ flux-flux
diagram as shown in Fig.~\ref{fig_fvg_pg0003}.
All data are corrected for galactic foreground extinction 
(Tab.~\ref{table1}).
The fact that PG0003+199 was much brighter in 1979 (McAlary et al. 1983) 
than in 2009-2011 allows us to achieve a 
linear least-squares fit to the data points with sufficient accuracy.  
The fit yields an AGN slope $\Gamma_{BV} = 1.20 \pm 0.04$. 
This slope is at the steep end, but within the range determined by 
Winkler et al. (1992) for other type-1 AGN. 

To check the slope of PG0003+199 further, 
we searched in the NASA Extragalactic Database (NED) 
and found two more $B$ and $V$ 
data points taken with comparable apertures,
one with 15$\farcs$7 by McAlary et al. (1983) 
and one with 14$\farcs$0 by Elvis et al. (1994). 
These (extinction corrected) data points 
are displaced to the right of the 7$\farcs$8 slope in 
Fig.~\ref{fig_fvg_pg0003}, because they contain 
additional host emission from the extended disk which is
redder than the AGN emission.  
These two data points yield a slope $\Gamma_{BV} = 1.25 \pm 0.04$.
The consistency of $\Gamma_{BV}$ derived from two independent data
sets corrobborates the steep AGN slope of PG0003+199.

The range of host slopes ($0.4 < \Gamma_{BV}^{host} < 0.53$) 
is taken from Sakata et al. (2010).
The intersection area of the AGN and host slopes 
defines the host flux. 
Averaging over the intersection area yields a mean host flux of 
2.16 $\pm$ 0.29 mJy in $B$ and 4.56 $\pm$ 0.30 mJy in $V$. 

During our monitoring campaign, the total $B$ fluxes
lie in the range between 5.69 mJy and 6.78 mJy 
with a mean of 6.24 $\pm$ 0.31 mJy. 
To derive the contemporaneous 5100\AA~ AGN luminosity,
which is widely used in reverberation studies, we tried to   
extrapolate the host-subtracted $B$ fluxes to restframe 5100\AA~ fluxes, 
using a power-law spectral shape ($F_{\nu} \propto \nu^{\alpha}$) 
with $\alpha$ being constrained by $\Gamma_{BV}$. 
But this extrapolation results in large uncertainties, 
because $\alpha$ depends sensitively on errors in $\Gamma_{BV}$. 
Therefore we applied the following procedure,
inferring in a first step the $V$ flux range.

During our monitoring campaign, the $V$ fluxes should lie in 
the intersection area of the $B$ flux range with the 
AGN slope (blue shaded area in Fig.~\ref{fig_fvg_pg0003}). 
The fact that the data point from June 2011 lies in this area 
supports our suggestion.
From this area we infer the $V$ fluxes during the monitoring campaign, 
yielding a mean  $fV$ = 7.95 $\pm$ 0.31 mJy.

Then, during our monitoring campaign the host-subtracted AGN fluxes are
in the range 4.08 $\pm$ 0.43 mJy in $B$ and 3.39 $\pm$ 0.40 mJy in $V$. 
From this range we interpolate 
the host-subtracted AGN flux of PG0003+199 at restframe 5100\AA~ 
$F_{5100\AA~}$ = 3.52 $\pm$ 0.41 mJy,
adopting for the interpolation 
that the AGN has a power law spectral energy distribution 
($F_{\nu} \propto \nu^{\alpha}$). 
To determine the errors on $F_{5100\AA~}$, 
we interpolated $F_{5100\AA~}$ between $fB+\sigma$ and $fV+\sigma$ 
as well as between $fB-\sigma$ and $fV-\sigma$. 
At the distance of 112.6 Mpc 
this yields 
a host-subtracted AGN luminosity 
$L_{5100\AA~} = 3.13 \pm 0.36 \times 10^{43} erg/s$.
Note that the 12\% uncertainty includes  the 
measurement errors, the uncertainty of the AGN and host slopes, 
and the AGN variations. 

The AGN luminosity during our campaign derived with the FVG method 
is about a factor two smaller than the value 
$L_{5100\AA~} = 6.03 \pm 0.03 \cdot 10^{43} erg/s$ 
derived with host galaxy modelling
by Bentz et al. (2009) for the data of the spectroscopic  
monitoring campaign (Peterson et al. 1998a). 
To understand this difference, we compared the 
numbers given in Bentz et al. (their Tables 7-9) with ours.   
Our aperture area ($7\farcs8$ in diameter) is only 25\% 
larger than that of
Peterson's campaign ($5\arcsec \times 7\farcs6$), 
and the additional flux from the outer region is negligible 
(Fig.~3 in Bentz et al. 2009). 
The total observed (not extinction corrected) 
fluxes $F_{5100\AA~}$ are similar, being 
$7.83 \times 10^{-15} erg s^{-1} cm^{-2} \AA^{-1}$ 
for our campaign and  
$7.68 - 8.81 \times 10^{-15} erg s^{-1} cm^{-2} \AA^{-1}$ 
for Peterson's campaign.
The main difference, however, lies in the estimate of the 
host contribution. 
Already in the early stage of the HST image decomposition 
with GALFIT, Bentz et al.'s host flux 
($fV_{host} \sim F550M = 1.88 \times 10^{-15} erg s^{-1} cm^{-2} \AA^{-1}$) 
corresponds after extinction correction to $2.12 mJy$, which is  
more than a factor two smaller than our $fV_{host} = 4.56 mJy$. 

As illustrated in Fig.~\ref{fig_fvg_pg0003}, it is hard to bring 
the small host flux from GALFIT modelling into agreement with the available 
FVG data. 
Firstly, we consider whether the discrepancy may be caused by failures
of the FVG method. We find:
1) Altering the assumptions on the host slope does not bring the host
contributions into agreement.
2) Although the AGN slope is determined by only a few data points, the
large flux range, which allows for a good slope determination, and the 
consistency between the AGN slope determined with apertures of 
7$\farcs$8 and $\sim$15$\arcsec$ argues in favour of a steep AGN
slope. 
3) The correction for galactic foreground extinction may introduce
errors in the AGN
slope. However, the foreground extinction
is relatively small (Table~\ref{table1}) and even without correcting 
for foreground extinction
the AGN slope remains steep ($\Gamma_{BV} =1.16 \pm 0.05$)  resulting in 
$fV_{host} = 4.2 \pm 0.3 mJy$, hence about twice the GALFIT host value.
4) If the host flux from GALFIT modelling were correct, then at low
luminosity the AGN slope (of PG0003+199) would show a strong curvature towards 
redder colours, in contradiction to the results by Sakata et
al. (2010) where the host galaxy lies on the linear extension of the AGN
slope. These four arguments lead us to conclude that the discrepancy
between GALFIT and FVG host fluxes is not caused by errors in the FVG
method alone. 
On the other hand, looking for potential error sources in the GALFIT
method, 
it could be that a large fraction of  
the bulge and/or nuclear starbursts of PG0003+199 
is unresolved on the HST images and therefore 
underestimated by the GALFIT modelling.  
Future data may clarify this issue.

\subsection{Results on Ark\,120}

\subsubsection{Implications from the spectrum}

\begin{figure}
  \includegraphics[angle=90,width=\columnwidth]{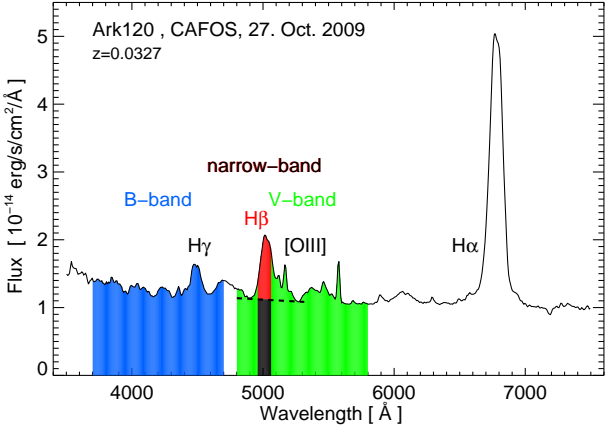}
  \caption{
    Observed spectrum of Ark\,120.
    The $B$-band (blue shaded) traces the AGN (and host) continuum and
    a small ($< 5$\%) contribution by H$\gamma$. The $V$-band (green
    shaded) traces the continuum and a small ($< 15$\%) contribution
    by H$\beta$. The narrow-band traces the continuum (black shaded)
    and H$\beta$ (red shaded), each component contributing to about
    50\% to the band pass. The red wing of H$\beta$ extending beyond 
    [OIII] is neither seen in 
    H$\alpha$ nor in H$\gamma$ and could be due to Fe lines 
    (Korista 1992).
      For the effective transmission curves of $B$ and $V$ see 
      Figure~\ref{fig_transmission}.
  }
  \label{fig_spectrum_ark120}
\end{figure}

Ark120 
lies at redshift $z = 0.0327$ so that the H$\beta$ line falls into the
NB\,$5007 \pm 30$\,\AA\, filter. Compared to the case of PG0003+199,
the observations of Ark\,120 are more challenging: Firstly, compared
to H$\alpha$, the H$\beta$ line is fainter contributing only to
$\sim$50\% of the flux in the NB band pass
(Fig.~\ref{fig_spectrum_ark120}). In order to measure the continuum
variations underneath the H$\beta$ line, light curves using
neighbouring off-line intermediate-bands would be ideal but such bands
were not available. While the $V$-band covers H$\beta$ and the Fe
complex, 
it is dominated to 83\% $\pm$ 3\% by the AGN and host continuum 
(calculated using the transmission curve of filter and CCD shown in 
Fig.~\ref{fig_transmission})
and therefore we used the $V$-band, in addition to $B$-band and
NB\,5007\,\AA. The second challenge is that (after accounting for 
the spectral resolution) the H$\beta$ line is
broader than the narrow bandpass, so that the broadest line wings,
carrying about $10 -15$\% of the line flux, escape detection in the
NB. Despite of these handicaps, the results are promising. 

\subsubsection{Light curves and BLR size}

\begin{figure}
  \centering
  \includegraphics[angle=0,width=\columnwidth]{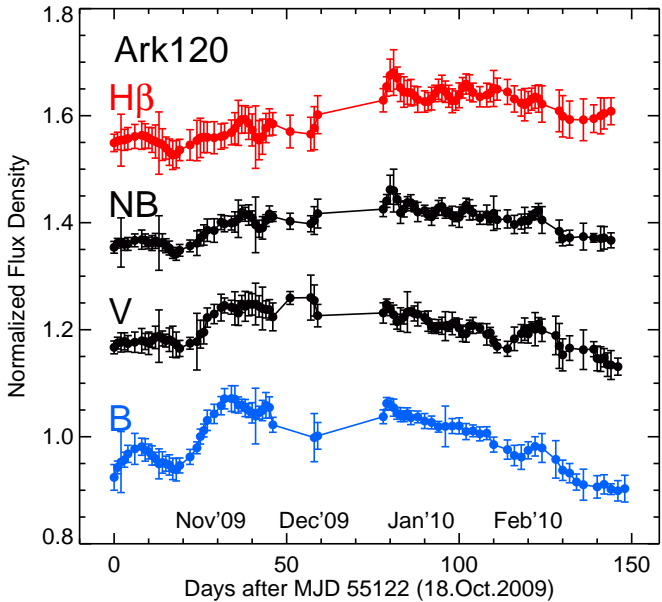}
  \caption{
      Light curves of Ark\, 120 between Oct. 2009 and March 2010. The
      $B$, $V$ and NB curves are as observed, the H$\beta$ curve is
      computed by subtracting a scaled $V$ curve from the NB curve
      (H$\beta$ = NB $-$ 0.5 $V$) and re-normalization (mean=1). For
      better visibility, the light curves are shifted against each
      other by $+0.2$. The December 2009 data are mostly lost due to
      strong wind preventing us from opening the dome. 
  }
  \label{fig_lightcurve_ark120}
\end{figure}

Fig.~\ref{fig_lightcurve_ark120} depicts the light curves of
Ark\,120. The dense time sampling clearly outlines the variations. The
$B$-band steeply increases during ten days November 2009 by about 17\% with a
sharp peak at end of November 2009; from January to March 2010 it
declines gradually by about 17\%. In November the $V$ and NB light curves are
similar to that of the $B$-band, but with smaller amplitude 
($V$ $\sim$ 10\%, NB $\sim$ 7\%). 
In contrast to the steep $B$ band flux increase, 
the NB flux increase is
stretched until January 2010 with an amplitude of 12-15\% between
begin of November and January. 
The possible influence of the missing data in December 2009 is
discussed in Section~\ref{section_discussion}.

The NB contains about 50\% continuum
and 50\% H$\beta$ (Fig.~\ref{fig_spectrum_ark120}). 
In order to remove the continuum, we computed a synthetic H$\beta$ light
curve by subtracting a scaled $V$ curve from the NB
curve: H$\beta$ = NB $-$ 0.5 $V$ ($V$ scaling factor = 0.5). 
Alternating the 
$V$ scaling factors between 0.4 and 0.6 yields
similar results. The synthetic H$\beta$ curve still has a remaining
variability pattern
in Oct-Nov which we suggest to arise from the relicts of the 
continuum. However, the important result is that compared to November
the $\sim$10\% higher flux level in January becomes more pronounced. 
 While the subtraction of a scaled $V$ curve from the NB curve 
turns out to be a 
successful approach to recover a pure H$\beta$ light curve, 
we discuss potential refinements 
in Section~\ref{section_discussion}.

\begin{figure}
  \centering
  \includegraphics[angle=90,width=\columnwidth,clip=true]{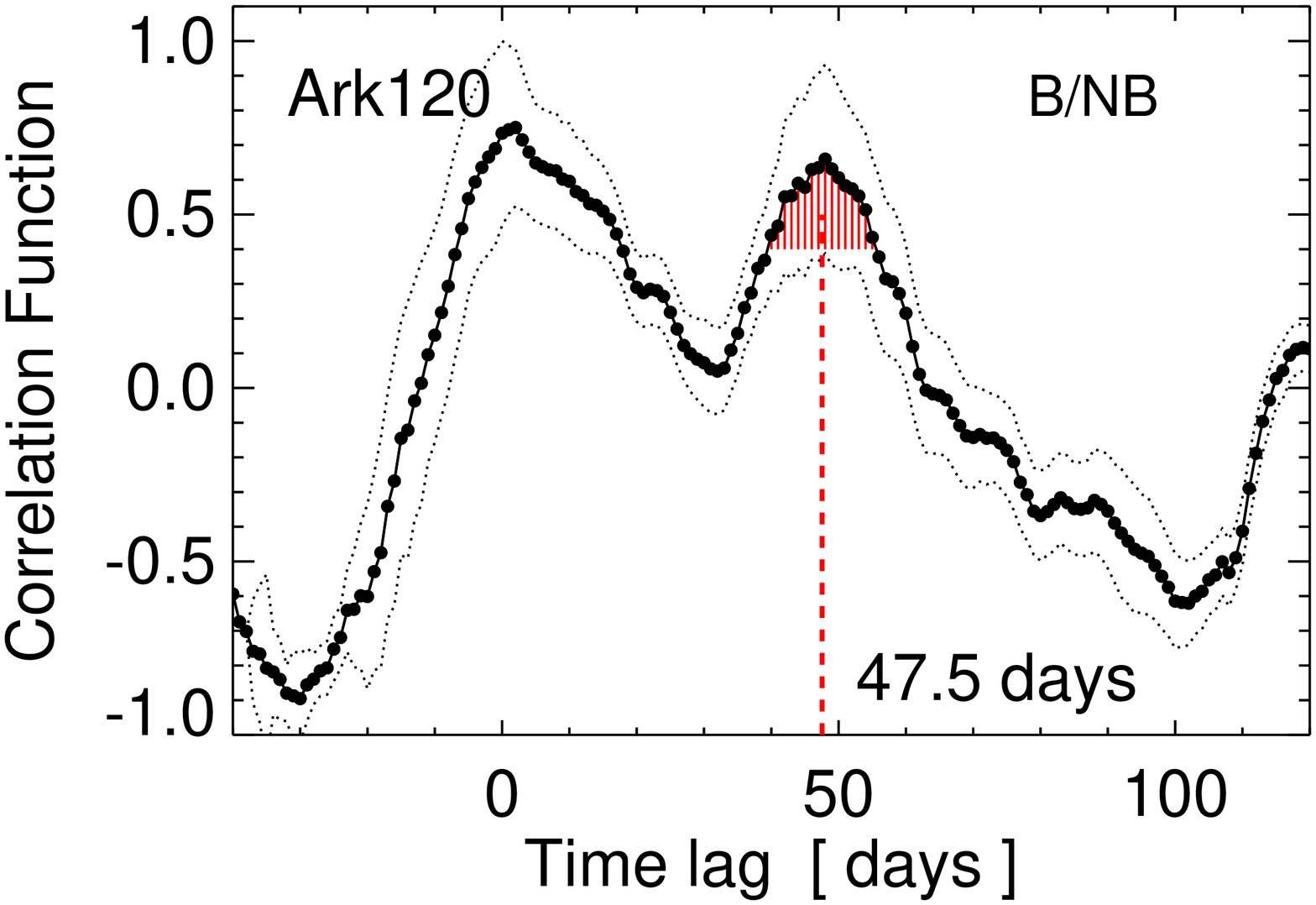}
  \includegraphics[angle=90,width=\columnwidth,clip=true]{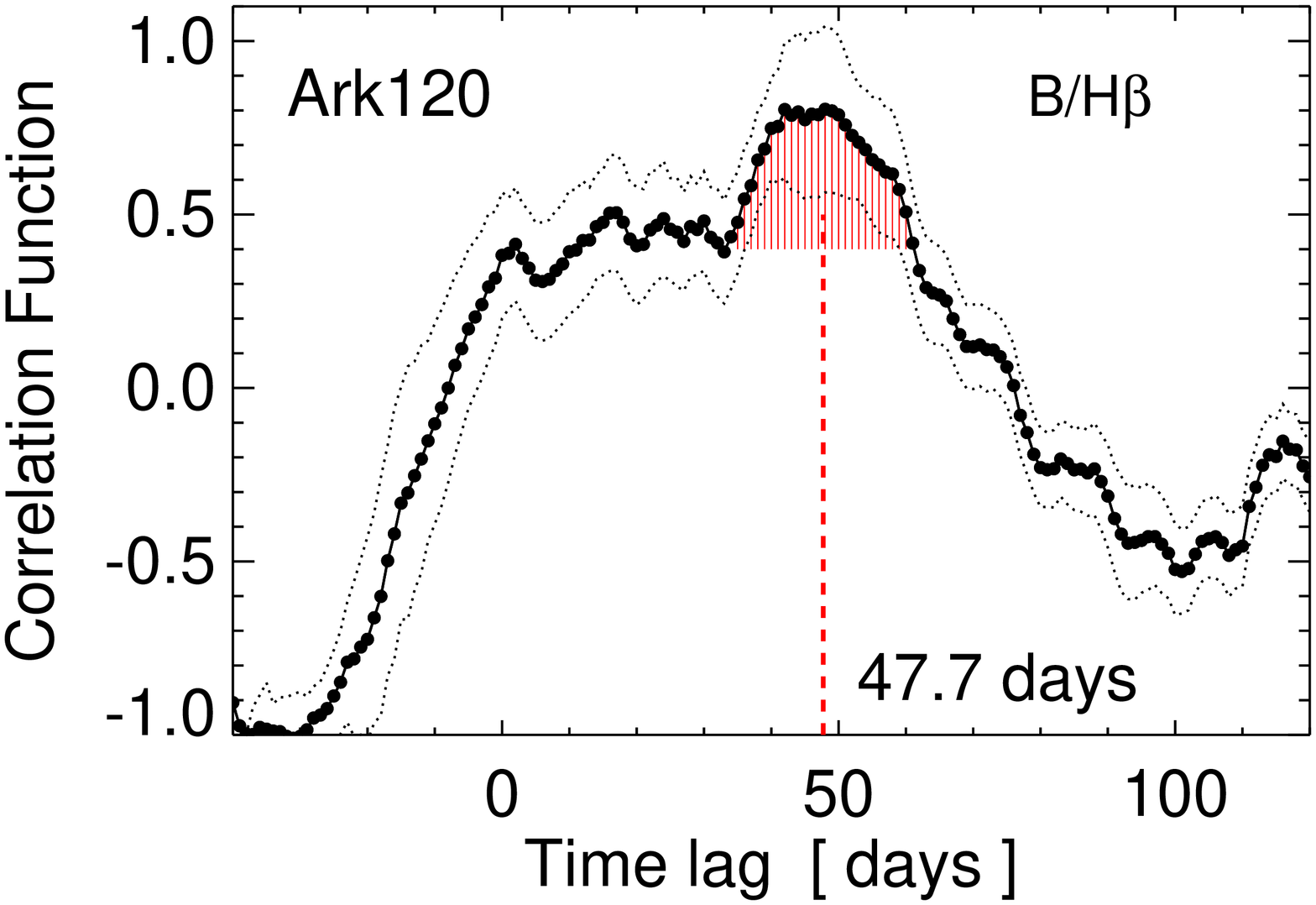}
  \caption{
    Cross correlations of Ark\,120 light curves for $B$ and
    NB, and for $B$ and the synthetic H$\beta$ (= NB $-$ 0.5
    $V$). 
      The dotted lines indicate the error range  ($\pm 1\sigma$) around the 
      cross correlation.
    The red shaded area marks the range used to calculate the
    lag $\tau_{\rm cent}$ (vertical dashed line). 
  }
  \label{fig_dcf_ark120}
\end{figure}

Fig.~\ref{fig_dcf_ark120} shows the cross correlation
results applying in a first step the DCF to the entire light curves.
The DCF of $B$ and NB
exhibits two distinct major peaks, one at a lag of 47.5 days and
 one at a lag of about zero days. 
Because the cross correlation of $B$ and $V$ yields 
a lag of 1.2$^{+1.6}_{-2.6}$ days consistent with lag zero, we
conclude that
the lag zero peak in the DCF of $B$ and NB comes from 
the continuum contribution contained in the NB. 
This contribution appears here quasi in autocorrelation, while the 
lag of 47.5 days is due to the line echo. 
This is further
corroborated by the DCF of the $B$ and synthetic $H\beta$ curves,
where the peak at lag 47.7 days dominates, while the continuum caused
contribution at lag zero is much reduced. 
This shows that the subtraction of a scaled $V$ light curve from the 
narrow-band light curve, in fact, largely removes the interferring
continuum emission, allowing us to  measure 
the lag of the emission line. 

\begin{figure}
  \centering
  \includegraphics[angle=90,height=3.5cm,clip=true]{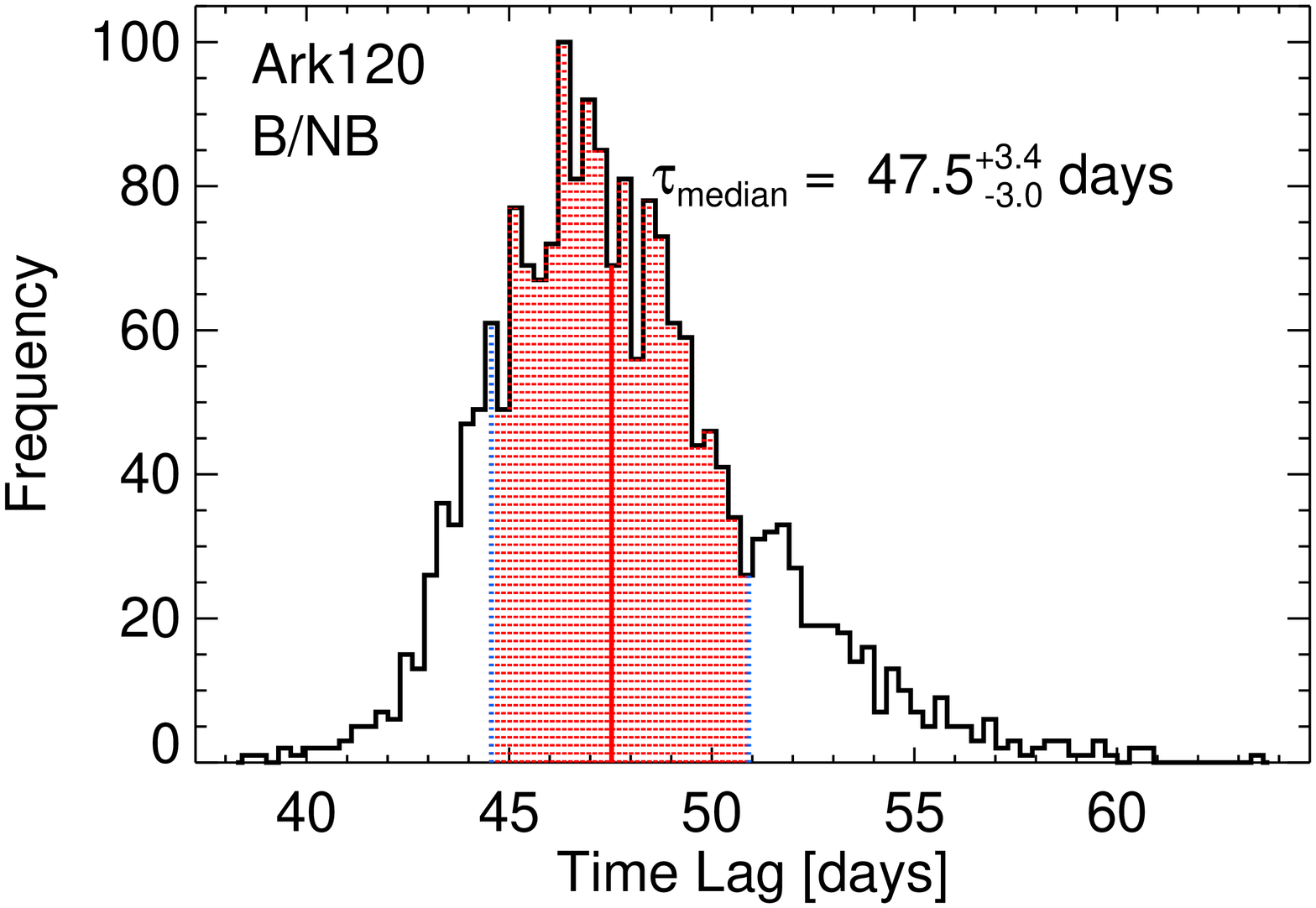}
  \includegraphics[angle=90,height=3.5cm,clip=true]{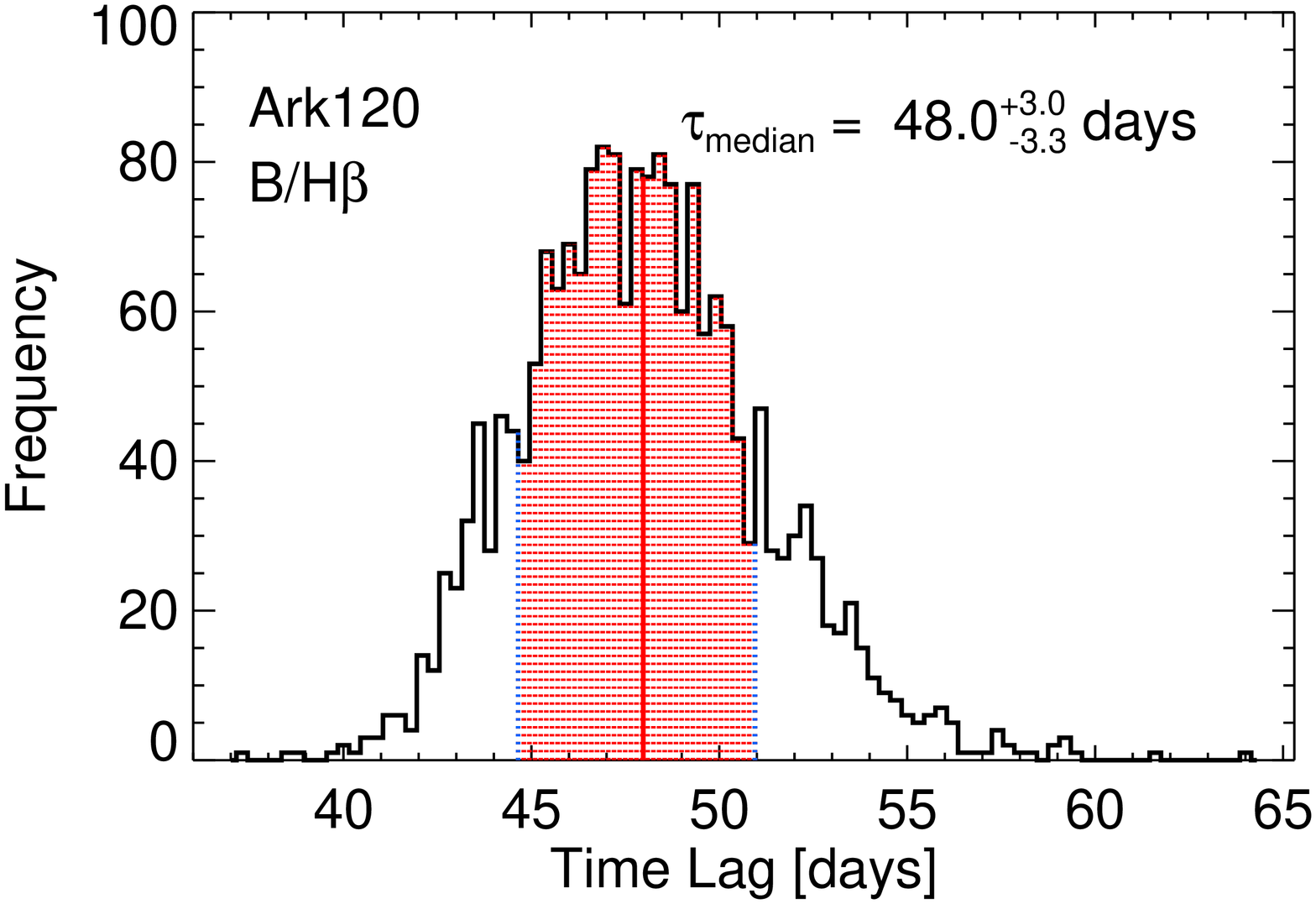}
  \caption{
    FR/RSS cross correlation results of Ark\,120. {\it left}: $B$ and
    NB, {\it right:} $B$ and synthetic H$\beta$.
    Each histogram shows the distribution of
    $\tau_{cent}$ obtained by cross correlating  
    2000 flux randomised and randomly selected
    subset light curves. 
    The median of this distribution is taken as 
    lag $\tau_{cent}$.   
    The red shaded area marks the 68\%  confidence range used to calculate the
    errors of $\tau_{\rm cent}$. 
  }
  \label{fig_hist_ark120}
\end{figure}

To determine the lag uncertainties, we applied the FR/RSS method,
again creating 2000 randomly selected subset light curves, 
each containing 63\% of the original data points,
and randomly altering the flux value of each data point consistent 
with its (normal-distributed) measurement error.
Then we cross correlated the 2000 pairs of subset light curves and
computed the centroid $\tau_{cent}$. As shown in Fig. \ref{fig_hist_ark120}, 
the resulting median lag is $\tau_{cent}$ = 47.5 $^{+3.4}_{-3.0}$ days
for the DCF of $B$ and NB,  and $\tau_{cent}$ = 48.0 $^{+3.0}_{-3.3}$ days 
for the DCF of $B$ and synthetic H$\beta$.
For short, the lag uncertainty is about 7\%. 
Correcting 48.0 days for the time dilation factor (1.0327) 
we obtain a rest frame 
lag of 46.5 $\pm$ 3.25 days, 
consistent with the lag found by
spectroscopic monitoring ($47 \pm 10$ days by 
Peterson et al. 1998a, 2004, 
and in the range 34-54 days by Doroshenko et al. 1999). 

\subsubsection{Black hole mass}

From our single epoch spectrum we determined an intrinsic H$\beta$ line 
dispersion $\sigma = 1950$\,km/s, 
in agreement with the value by Peterson et al. 
(2004). To get rid of the red wing of the H$\beta$ line, we mirrored 
the blue side of the H$\beta$ line profile to the red, 
and calculated the
line dispersion of this synthetic profile. 
The shape and the dispersion of this profile are 
consistent with what we derived for 
H$\alpha$ ($\sigma \sim 1970$\,km/s) 
and H$\gamma$ ($\sigma \sim 1920$\,km/s). 
Combining  line 
dispersion  (with 25\% uncertainty adopted) 
and the time delay (46.5 $\pm$ 3.25 days) yields a virial
$M_{\rm BH}  =  34 \pm 12 \times 10^{6}$\,M$_\odot$, consistent with the
value of $35 \pm 8 \times 10^{6}$\,M$_\odot$ derived via spectroscopic
reverberation mapping by Peterson et al. (2004).
 
Already in the 1970s and 1980s numerous studies revealed that 
the H$\beta$ line profile of Ark120 is strongly variable and 
exhibits two peaks in the rms spectra and a prominent red wing, 
which is neither present in H$\alpha$ nor in 
H$\gamma$ (Fig.~\ref{fig_spectrum_ark120}). 
As discussed by Doroshenko et al. (1999), 
the BLR of Ark120 may show accretion inflows in addition to the 
virialised motion, and therefore any black hole mass 
estimates should be considered with some reservation.

\subsubsection{Host-subtracted AGN luminosity}
\label{section_ark120_fvg}

\begin{figure}
  \centering
  \includegraphics[angle=0,width=\columnwidth,clip=true]{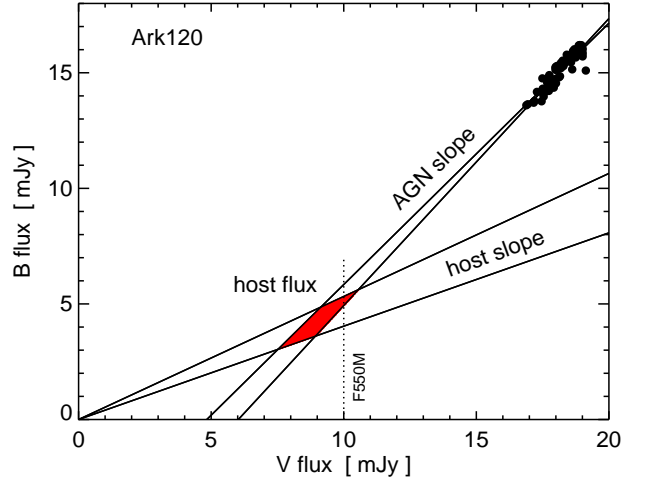}
  \caption{
    $B$ versus $V$ fluxes of Ark\,120, measured in a 7.5$\arcsec$
    aperture. The fluxes for each night (fat dots) follow a linear
    gradient yielding the range for the AGN slope. 
    As host slope we adopted the range found by Sakata et al. (2010). 
    The intersection of the two slopes (red shaded area) defines the 
    host flux. 
    The vertical dotted line marks the host flux found 
    in an aperture of $5\arcsec \times 7\farcs6$ 
    via modelling of HST F550M images by Bentz et al. (2009).  
    All data are corrected for galactic foreground extinction. 
  }
  \label{fig_fvg_ark120}
\end{figure}

To separate AGN and host luminosity contributions, we applied the flux
variation gradient method  (Choloniewski 1981). 
Fig.~\ref{fig_fvg_ark120} shows the $B$ versus $V$ fluxes of
Ark\,120, 
corrected for galactic foreground extinction (Schlegel et al. 1998).
Table \ref{table2} summarizes the results. 
A linear least-squares fit to the flux variations yields 
$\Gamma_{BV} = 1.19 \pm 0.06$, 
consistent with $\Gamma_{BV} = 1.20 \pm 0.10$
found by Winkler et al. (1992, 
using 20$\arcsec$ aperture and adopting $E_{B-V}=0.1$)
and with  $\Gamma_{BV} = 1.22 \pm 0.02$ 
found by Doroshenko et al. (2008, after correcting the 
values in their Table~3 for extinction; 
note that due to a  typo 
the value $V = 14.27 mag$ in this Table should
be $V = 14.72 mag$, in order to be consistent with the plotted values
in their Fig.~4).
The range of host slopes is taken from Sakata et al. (2010).
The intersection area of the AGN and host slopes 
defines the host contribution. 
Averaging over the intersection area yields a mean host flux of 
$fB$ = 4.26 $\pm$ 0.59 mJy and $fV$ = 9.02 $\pm$ 0.62 mJy. 
Our host fluxes (in a 7$\farcs$5 aperture) 
are consistent, albeit a bit high, when compared with the values 
$fB \approx 4.5 mJy$ and $fV \approx 9 mJy$  (in a 15$\arcsec$ aperture) 
from a FVG analysis of $UBVRI$ data (Doroshenko et al. 2008).
Subtracting our host fluxes from the total fluxes, 
we obtain the host-subtracted $B$ and $V$ fluxes and 
then by power-law interpolation 
the rest-frame AGN flux $f_{5100\AA~} = 9.41 \pm 0.86 ~mJy$. 
At the distance of 141.8 Mpc, this yields a
host-subtracted AGN luminosity 
$L_{5100\AA~}=13.27 \pm 1.21 \times 10^{43} erg/s$. 
The uncertainty of the AGN luminosity is less than 10\%.

The AGN luminosity during our campaign as 
derived with the FVG method 
is about 50\% higher than the mean value 
$L_{5100\AA~} = 8.47 \pm 0.81 \times 10^{43} erg/s$ derived  
with host galaxy modelling
by Bentz et al. (2009) for the data of the spectroscopic 
monitoring campaign (Peterson et al. 1998a). 
A detailed comparison of the parameters yields 
that our aperture area
($7\farcs5$ in diameter) is only 16\% larger than 
that of Peterson's campaign ($5\arcsec \times 7\farcs6$), 
and the  $V$ band host 
fluxes agree within the uncertainties 
as shown in Fig.~\ref{fig_fvg_ark120}
(extinction corrected: $fV \approx 9 mJy$ from our data and 
$F550M \approx 10 mJy$ from the HST/spectroscopic data base).
However, the total observed (not extinction corrected) 
fluxes $F_{5100\AA~}$ differ significantly, 
being $13.4 \times 10^{-15} erg s^{-1} cm^{-2} \AA^{-1}$ for our
campaign and lying in the range  
$7.82 - 10.37 \times 10^{-15} erg s^{-1} cm^{-2} \AA^{-1}$ 
during Peterson's eight years campaign, hence  
are 30\% -- 70\% larger at our campaign. 
This leads us to conclude that the AGN luminosity is  
$\sim$50\% higher during our campaign.

\section{Discussion}
\label{section_discussion}

The tests on PG0003+199 and Ark120 demonstrate
the feasibility of photometric
reverberation mapping to measure the BLR size. 
The case of Ark120 demonstrates that the continuum contribution 
($\sim$50\%) to the narrow-band NB can successfully be 
estimated and corrected (H$\beta$ = NB -- 0.5~$V$), 
while the uncorrected cross correlation ($B$/NB) shows 
two distinct peaks, one peak from the autocorrelation of the
continuum and one peak from the emission line 
at lag $\tau_{cent}$ consistent with the literature data. 
In the case of Ark120, it turns out that 
the host contribution to the $V$ band is 50\% and 
the AGN+host continuum contribution to the NB is about 50\%, too. 
Therefore, due to this coincidence, 
the $V$ scaling factor of 0.5 works well. 
The general
procedure, to recover a pure H$\beta$ light curve,  
depends on the AGN/host contrast $C_{AGN/host}$ in the filters used 
and the AGN/line ratio $C_{AGN/line}$ in the NB.
Thus, a proper  treatment of the relevant 
filter light curves would be: 
\begin{itemize}
\item[$\bullet$] 
Remove the constant host emission from the light curves:
To this end, estimate the host contribution  in each filter, 
for instance via FVG analysis.
Subtract the host contribution from the light curves.
Now the (renormalized) light curves 
measure the AGN continuum 
and in the NB also the line emission.
\item[$\bullet$]
Extract the line light curve from the NB light curve:
To this end, 
estimate the AGN continuum contribution to the NB, 
for instance from a
spectrum also taking aperture differences into account.
Then, using a suitable filter next to the NB, e.g. $V$ for H$\beta$,
subtract the AGN continuum light curve from the NB light curve.
\end{itemize}
This procedure requires good signal/noise data.
Applying the procedure on Ark120 for a range of 
$C_{AGN/host}$ and $C_{AGN/line}$ parameters, 
we obtained cross correlation results in agreement with those of 
the straight forward method (H$\beta$ = NB -- 0.5~$V$).
This indicates that the lag determination is quite robust 
against parameter choice to remove the host and AGN continuum.  
A more detailed simulation of these issues would be interesting, 
but is beyond the scope of this paper.  

We now discuss the accuracies achieved for the two objects,  
the importance of good time sampling, and the improvement 
potential for the $R_{\rm BLR} - L$ relationship.

The BLR sizes determined by photometric monitoring 
agree within a few percent with
those from spectroscopic reverberation mapping.
The $R_{BLR}$ measurement accuracy we derived from the FR/RSS
method
is 12\% for the short light curve of PG0003+199 and 7\% for the 
longer monitored Ark120, while the uncertainty of the
black hole mass ($\sim$35\%) is mainly due to the adopted 25\% uncertainty 
of the velocity dispersion from single epoch spectra. 

Our $R_{BLR}$ measurement accuracy appears exceptionally high
compared with typical $R_{BLR}$ errors ($\sim$30\%) reported 
so far from spectroscopic reverberation data. 
Adopting that the determination of the emission line lag 
via spectroscopic monitoring is 
more precise than via photometric monitoring, 
the difference in the error ranges
may be explained by two effects.
Firstly, compared to our photometric campaign, most of 
the available spectroscopic campaigns were performed with lower
sampling rate, but over longer time spans. Thus they catch several
variability events and implicitly determine an average lag. 
Then the reported errors include not only the formal measurement
errors, but also the possible scatter of the intrinsic lags.
On the other hand, a low sampling rate might limit the potential to reach
small errors. Then our high $R_{BLR}$ measurement 
accuracy might be favoured by the high 
sampling rate (median 2 days) of the light curves. 

\begin{figure}
  \centering
  \includegraphics[angle=0,width=\columnwidth,clip=true]{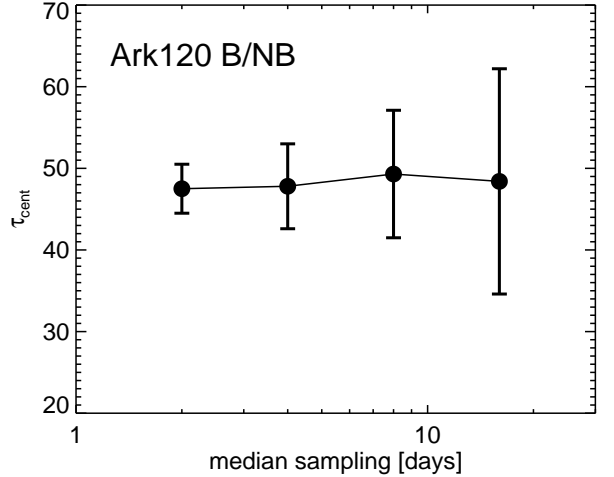}
  \caption{
    Dependence of  $\tau_{cent}$ and its errors from the median
    sampling. 
  }
  \label{fig_ark120_tau_error}
\end{figure}

To check the influence of the sampling rate on the errors, 
we performed the following test. 
From the entire light curves of Ark120 we created three regular 
subsets with a median sampling of 4, 8, and 16 days and 
determined for each subset $\tau_{cent}$ and its errors
using the FR/RSS analysis. 
It turns out that poorer sampling results in substantially 
increased errors.
Even worse, for the poorest data set (sampled with median 16 days) 
$\tau_{cent}$ could not be recovered. The reason is that 
the two cross correlation peaks seen in Fig.~\ref{fig_dcf_ark120}
were not anymore separated. Therefore, 
in order to quantify the error dependence from the sampling, 
we applied a constraint in FR/RSS analysis. 
We excluded those $\tau_{cent}$ values outside the range 30--70 days. 
This enables us to determine   
$\tau_{cent}$ even in the poorly sampled data set.  
The result of this procedure is shown in Fig.~\ref{fig_ark120_tau_error}.
The test confirms that the sampling rate plays a crucial role to
achieve small errors. 
More important, if the continuum contributes 
significantly to the narrow-band, 
then photometric reverberation mapping requires well sampled light
curves, because otherwise the cross correlation may fail to disentangle 
the line echo from the autocorrelated continuum contribution. 

\begin{figure}
  \includegraphics[angle=90,width=\columnwidth]{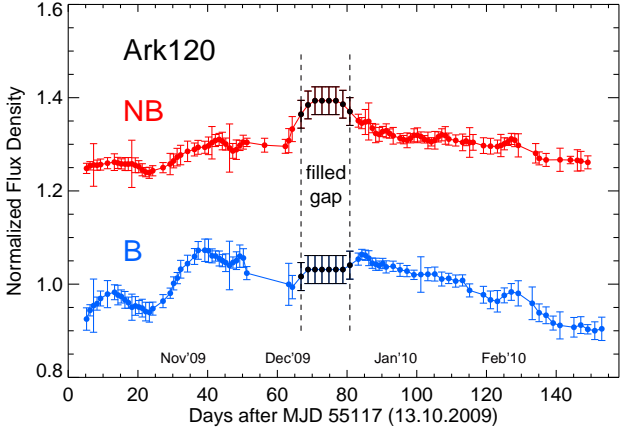}
  \caption{
     Artificial $B$ and NB light curves of Ark120, 
     with additional data points (black dots) filling the
     observational gap of 18 days end of December 2009.
     The data points are chosen so that they 
     mimic a strong line echo during the gap.
  }
  \label{fig_ark120_lc_artificial}
\end{figure}

\begin{figure}
  \includegraphics[angle=90,width=\columnwidth,clip=true]{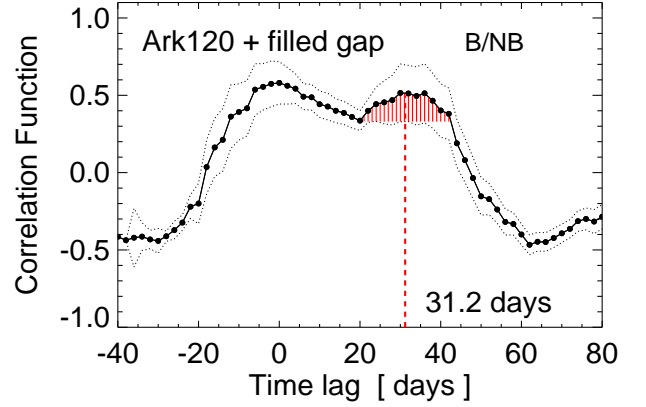}
  \includegraphics[angle=90,width=\columnwidth,clip=true]{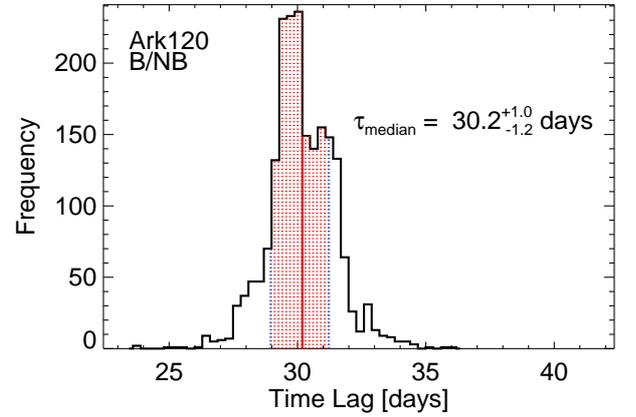}
  \caption{{\it Top:} Cross correlation of the complemented 
    $B$ and NB light curves of Ark\,120 
    shown in Fig.~\ref{fig_ark120_lc_artificial}. 
      The dotted lines indicate the error range  ($\pm 1\sigma$) around the 
      cross correlation.
   {\it Bottom:} Distribution and 68\% confidence range of
    $\tau_{cent}$, obtained by  FR/RSS analysis.
  }
  \label{fig_hist_ark120_artificial}
\end{figure}

We caution that the formal errors from the FR/RSS analysis do not
account for potential systematic errors
caused, for instance, by gaps in the observed light curves.
We illustrate this with the light curves of Ark120.
Despite the dense median sampling of the light curves by 2 days, 
the light curves of Ark120 have a gap of 18 days 
in December 2009 (Fig.~\ref{fig_lightcurve_ark120}). 
While the continuum brightening occurred in November,  
the line brightening, as far as we can see it, is delayed to January 2010, 
yielding the measured lag of $\tau_{cent} \sim 48 days$.
However, because of the observational 
gap we do not know whether the line brightening 
occurred already in December 2009.
To test the effect, if there were a strong line echo in December 2009, 
we took the $B$ and NB light curves and filled the gap with 
the artificial bump in the NB light curve as shown 
in Fig.~\ref{fig_ark120_lc_artificial}.
The bump may be somewhat strong, 
but it was chosen to better  
illustrates the potential effects here. 
One can see already in the light curves 
that the lag between the $B$ band increase in November and
the NB increase in December is about 30 days.
The cross correlation of these artificial light curves
shows two major peaks, one from the autocorrelated continuum at lag
zero and one from the line at lag 31 days
(Fig.~\ref{fig_hist_ark120_artificial}).
The FR/RSS analysis yields $\tau_{cent} = 30.2^{+1.0}_{-1.2}$ days. 
The difference between this lag 
and $\tau_{cent} = 48$ days derived from the original light curves
is 18 days which is just about the size of the gap.
From this example 
we conclude that in case of a large observational gap 
one should take a potential systematic error of $\tau_{cent}$ into
account, which is about the size of the gap. 
In this example the light curve 
is short in the sense that it covers only one
pronounced continnum event and one subsequent line echo.
Therefore the missing information during the gap cannot 
be filled. But in long lasting light 
curves which cover several variability events 
one may expect that on average the effective gaps are smaller.
These considerations about systematic errors hold
for both photometric and spectroscopic reverberation data.
Therefore, they 
do 
not question the successful applicability of photometric
reverberation mapping.

Reverberation mapping, in general, relies on the time
resolution, and therefore it is 
vital that the light curves are sufficiently well sampled. 
Reminding Nyquist's Theorem, the sampling rate SR of the feature 
of interest, i.e. the lag, must be SR = $t_{lag}/t_{sampling} > 2.2$. 
In the observational practice, data points are lost due to poor 
weather or technical problems, hence one should aim at SR $>$ 3.
So far, however, many BLR sizes have been derived from sparsely sampled 
light curves (SR $<$ 3) with a limited potential to reach better 
than 30\% accuracy. It could be that for undersampled 
data any quoted uncertainties of 30\% are too 
optimistic. 
For example, the recent spectroscopic reverberation campaign 
with improved daily sampling on NGC\,4051 revealed that its BLR 
size is actually a factor $\sim$3 smaller than inferred 
from earlier undersampled data (Denney et al. 2009b). 
Therefore, we suggest that some more of the BLR sizes measured 
so far can be improved with well sampled reverberation data. 

Applying the flux variation gradient (FVG) method to $B$ and $V$ band 
data and adopting a host colour range determined for other local AGN,
we find that for both objects PG0003+199 and Ark120 the host-subtracted AGN 
luminosity  $L_{5100\AA~}$ differs significantly from previous 
measurements. 
While for PG0003+199 we find a stronger host contribution leading to
50\% lower AGN luminosity, for Ark120 our host
estimate is consistent with others but the AGN
 was in a 50\% brighter state during our campaign.
The AGN/host luminosity contrast of both objects is relatively low, 
$L_{AGN}/L_{host} \sim 1$ for the $7\arcsec - 8\arcsec$ apertures used. 
The uncertainties of $L_{5100\AA~}$ are 
9\% and 12\% in Ark120 and PG0003+199.
These uncertainties include the measurement errors, 
the AGN variations and the uncertainty of the host flux. 
For $L_{AGN}/L_{host} \sim 1$, the errors are dominated by 
the uncertainty of the host flux.  
It depends on the error of the AGN slope and 
the range adopted for the host slope.
If improved image quality allows us to use smaller photometric apertures
containing less host flux,   
the uncertainty caused by the range of host slopes may be reduced.
Also, applying the FVG method to more than one filter pair, for instance
from $UBVRI$ monitoring data, allows one to construct several 
independent AGN slopes and thus to determine the host contribution in
a consistent manner, largely independent of the adopted range of host
slopes.  
For powerful AGN with $L_{AGN}/L_{host} > 2$ the uncertainty of the 
host contribution plays a minor role.

An important result from spectroscopic reverberation mapping is the
relationship between the BLR size and the nuclear luminosity 
$R_{\rm BLR} \propto L^{\alpha}$ with a predicted slope $\alpha$ = 0.5 (Netzer
1990; Netzer \& Marziani 2010). 
This relationship allows one to derive the virial
black hole mass for high redshift AGN 
from single epoch spectra by inferring $R_{\rm BLR}$ from $L$
(e.g. Vestergaard 2002; Netzer 2003, McLure \& Dunlop 2004).
While early observations indicated
$\alpha = 0.6 \pm 0.1$ (Kaspi et al. 2000, 2005), for H$\beta$ line
and 5100~\AA~ luminosity a recent analysis including host galaxy
subtraction yields $\alpha = 0.519 \pm 0.063$ (Bentz et
al. 2009). 
Note that the slope of the current $R_{\rm BLR} - L$ relationship 
depends on the adopted cosmology, because the more luminous sources 
are at higher redshift than the low-luminosity AGN.

\begin{figure}
  \centering
  \includegraphics[angle=90,width=\columnwidth,clip=true]{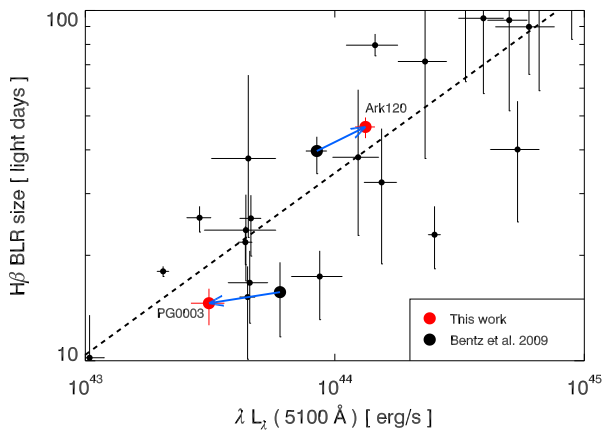}
  \caption{
    $R_{BLR}$ versus $L$, from Bentz et al. (2009). 
    Shown is a zoomed portion containing the two sources PG0003+199 and Ark120. 
    For PG0003+199 we used $R_{{\rm H}\beta} = R_{{\rm H}\alpha} / 1.37$ 
    (Sect.~\ref{section_pg0003_lc}).
    The new measurements shift both objects clearly outside the range of the previous errors 
    bars and about a factor 2 closer to the dashed 
    line which is the fit with a slope $\alpha = 0.519$ 
    obtained by Bentz et al.(2009).
  }
  \label{fig_r_l}
\end{figure}

The current $R_{\rm BLR} - L$ relationship exhibits  a large scatter 
over an order of magnitude 
in both $R_{\rm BLR}$ and  $L$, and many objects have large 
error bars (Fig.~\ref{fig_r_l}). 
The new photometric reverberation measurements  
shift the position of 
PG0003+199 and Ark120 in this diagram 
outside the quoted error range of the previous positions. 
Notably the  shift is larger in $L$ than in $R_{\rm BLR}$. 
The new positions are about 50\% closer to the relationship 
fitted by Bentz et al. 
(dashed line in Fig.~\ref{fig_r_l}). 
This suggests that well sampled reverberation data and improved 
host-free AGN luminosities may significantly reduce the scatter 
of the $R_{\rm BLR} - L$ relationship. 
Certainly, a true scatter may remain, 
simply because AGN are complicated objects and 
the $R_{BLR}$ determination depends
on the continuum variability pattern and 
the geometry of the BLR which produces the line echo. 
Future data will show how far the scatter can  be reduced. 

\section{Outlook}

While spectroscopic reverberation mapping is the only way to explore
the details of the innermost AGN structure and the geometry of the BLR
(e.g. Kollatschny 2003a, 2003b; Kollatschny \& Zetzl 2010), the
advantage of photometric reverberation mapping 
 with suitable filters is to efficiently measure BLR
sizes and host-subtracted luminosities 
for large AGN and quasar samples -- even with small
telescopes. Photometric broad- and narrow-band 
monitoring of a sample of 100 quasars ($V <
18~mag$) can be performed with 1-m telescopes in an equivalent of 3
years observing time. 

In the case of Ark120 the H$\beta$ line contributes to only 50\%
of the flux in the narrow-bandpass we used.
This suggests 
that photometric reverberation mapping of emission line lags works even
for broader bands, as long as the line contributes at least 50\%
to the bandpass. Also, it is worthwhile to test the method even for
weaker emission lines contributing less than 50\% to the bandpass. 
The upcoming Large Synoptic Survey Telescope (LSST) is equipped with
six broad band filters and will
discover thousands of variable AGN.
The H$\alpha$ line is shifted into the $i'$ and $z'$ bands
at $z \approx 0.16$ and $z \approx 0.38$,
respectively and may be strong enough, contributing about 50\% to these
broad bands. 
While an approximate photometric
redshift may be sufficient to determine the filter which contains the 
H$\alpha$ line,
it is desirable to take a spectrum of the AGN, in order to measure an 
accurate redshift, also for luminosity determination, 
and the line dispersion for black hole mass
estimates. In order  to determine the
H$\alpha$ line lag, the neighbouring filters, which are largely free
of line emission, may be used to interpolate and remove the continuum
variations underneath the H$\alpha$ line, 
in analogy to what we did for Ark120. 
Recently, an interesting statistical 
alternative to our method has been proposed by
Chelouche \& Daniel (2011), which is 
specifically designed for large samples of
AGN light curves expected to be obtained with the LSST. 
Consider AGN light curves in two bandpasses, curve  
X (tracing the continuum
largely free of emission lines) and curve Y 
(tracing the emission line with
underlying continuum): While our approach aimed to remove the
continuum contribution from curve Y by subtracting a scaled X curve, 
Chelouche \& Daniel propose to use the
light curves X and Y unchanged, and to subtract the autocorrelation of X 
from the cross correlation of X and Y, in order to determine the line
lag. 
Numerical simulations with synthetic AGN light curves 
and the treatment of four archival PG quasar light curves yield lags 
which are (in three of four cases) consistent with spectroscopic results. 
However, the reported lag uncertainties of individual AGN are large
($\sim$50\%) and 
only for averages of large ensembles of several hundred AGN the
obtained lags appear satisfying.   
Future optimisations, combining our approach with that of
Chelouche \& Daniel, would be intriguing.

Finally we outline a modification of the proposals by 
Elvis \& Karovska (2002) and Horne et al. (2003)
to determine quasar distances from reverberation data 
and thus to probe dark energy. 
The luminosity difference between the open Einstein -- de Sitter
cosmology ($\Omega_{\rm M} = 0.2, \Omega_{\Lambda}= 0$) and the
concordance cosmology ($\Omega_{M} = 0.3, \Omega_{\Lambda}= 0.7$) 
is 20\%--30\% at redshift $0.4<z<0.8$ (Riess et al. 1998).
From the $R_{\rm BLR} - L$ relationship we conclude that two quasar samples 
with the same $R_{\rm BLR}$ distribution, one sample at low redshift 
and one sample at high redshift, should have the same intrinsic 
$L$ distribution. 
Thus one may constrain the luminosity for different 
cosmologies by measuring 
the BLR size and the host-subtracted brightness 
of quasar samples at different redshift. 
If for objects matching in $R_{\rm BLR}$ the true 
scatter in $L$ can be reduced to about a factor 2 (= 200\%), 
the mean luminosity of 1000 quasars may be determined 
with a statistical accuracy of 200\%/$\sqrt 1000$ = 6\%, 
enabling a 3$\sigma$--5$\sigma$ detection of a cosmological 20\%--30\% 
luminosity difference. 

\begin{table*}
  \begin{center}
    \caption{ Source parameters. 
      Photometry is obtained using  apertures of 7$\farcs$8 for PG0003+199 and  7$\farcs$5 for Ark120.
      For the monitoring campaigns the photometry range is indicated by the double arrow.  
      The photometry errors of 2\%--4\% for individual measurements are determined 
      from the scatter caused by $\sim$10 different calibration stars from Landolt (2009). 
      To apply the flux variation gradient method, the photometry is corrected 
      for galactic foreground extinction and converted to mJy, 
      with results listed as  $fB_{ext~corr}$ and $fV_{ext~corr}$.
      Values used to correct for galactic extinction are:
      PG0003+199: $A_{B}$ = 0.153 mag, $A_{V}$ = 0.118 mag; 
      Ark120: $A_{V}$ = 0.554 mag, $A_{V}$ = 0.426 mag (Schlegel et al. 1998).  
      For the conversion from mag to mJy,
      zero mag fluxes of 4266.7 Jy in $B$ and 3836.3 Jy in $V$ are used. 
    }
    \begin{tabular}{l | r c | c | c c | c c}
      Object & Redshift & $D_{L}$ & Date                & $B_{obs}$                      & $V_{obs}$           & $fB_{ext~corr}$                  & $fV_{ext~corr}$     \\
             &          &  Mpc    &                    &  mag                           &   mag               &    mJy                         &   mJy             \\
      \hline                                                                                                                
             &          &      &                       &                                &                     &                                &                   \\
       PG0003+199& 0.0258&112.6& Aug. - Nov. 2009      & 14.65 $\leftrightarrow$ 14.84  &                     &  5.69 $\leftrightarrow$ 6.78   &                   \\
             &          &      & July 25$^{th}$, 2010   & 14.92 $\pm$ 0.037             & 14.45 $\pm$ 0.022    & 5.29 $\pm$ 0.18                &  7.07 $\pm$ 0.14  \\
             &          &      & June 26$^{th}$, 2011   & 14.78 $\pm$ 0.036             & 14.33 $\pm$ 0.033    & 6.01 $\pm$ 0.20                &  7.93 $\pm$ 0.24  \\
             &          &      & Aug 17$^{th}$, 1979 (McAlary et al. 1983)   & 13.71 $\pm$ 0.029             & 13.56 $\pm$ 0.021    &  16.10 $\pm$ 0.45              &  16.17 $\pm$ 0.31\\
             &          &      &                       &                                &                     &                                &                   \\
       Ark\,120& 0.0327 & 141.8& Oct. 2009 - March 2010& 14.11 $\leftrightarrow$ 14.30  &13.68 $\leftrightarrow$ 13.81& 13.59 $\leftrightarrow$ 16.20& 16.89 $\leftrightarrow$ 19.13     \\
             &          &      &                       &  $\pm$ 0.036                   &  $\pm$ 0.048        &                               &                   \\
             &          &      &                       &                                &                     &                                &                   \\
      
      \hline
    \end{tabular}
    \label{table1}
  \end{center}
\end{table*}

\begin{table*}
  \begin{center}
    \caption{  Parameters derived from the flux variation gradient method (FVG). 
      $\Gamma_{BV}$ denotes the FVG slope obtained by a linear least-squares fit 
      to the varying total $fB$ and $fV$ fluxes plotted in a $fB$ versus $fV$ diagram 
      (Figs. \ref{fig_fvg_pg0003} and \ref{fig_fvg_ark120}). 
      $fB_{total}$ and $fV_{total}$ refer to the mean of the total $fB_{ext~corr}$  
      and $fV_{ext~corr}$ flux ranges during our monitoring campaigns. 
      Note that $fV_{total}$ of PG0003+199 is not directly measured, 
      but inferred from $fB_{ext~corr}$ via the FVG slope (Fig. \ref{fig_fvg_pg0003}).
      $fB_{host}$ and $fV_{host}$ are derived from the intersection of the AGN and host slopes 
      (red shaded area in Figs. \ref{fig_fvg_pg0003} and \ref{fig_fvg_ark120}).
      The  range of host slopes is $0.40 < fB/fV < 0.53$, adopted from Sakata et al. (2010).
      $fB_{AGN} = fB_{total} - fB_{host}$ and $fV_{AGN} = fV_{total} - fV_{host}$, 
      with uncertainty range $\sigma_{AGN} = (\sigma_{total}^{2} + \sigma_{host}^{2})^{0.5}$. 
      $f_{AGN} ((1+z) 5100\AA)$ gives the (foreground extinction corrected) rest frame 5100\AA~flux, 
      interpolated from $fB_{AGN}$ and $fV_{AGN}$ for an AGN continuum with power law spectral shape ($F_{\nu} \propto \nu^{\alpha}$).
      $\lambda L_{\lambda,AGN} (5100\AA)$ is the monochromatic rest frame 5100\AA~luminosity. 
    }
    \begin{tabular}{l | c | c c c | c c c | c c | c}
      Object & $\Gamma_{BV}$ & $fB_{total}$ & $fB_{host}$ & $fB_{AGN}$ &$fV_{total}$ & $fV_{host}$ & $fV_{AGN}$ &\multicolumn{2}{c}{$f_{AGN}~((1+z) 5100\AA)$ ext corr} & $\lambda L_{\lambda,AGN} (5100\AA)$\\
             & & & & & & & & & & \\
             &  &    mJy   &    mJy   &    mJy  &   mJy    &    mJy     &    mJy     & mJy &  $10^{-15}~ erg~s^{-1} ~cm^{-2} ~\AA^{-1}$ &  $10^{43}~ erg~ s^{-1}$  \\
             & & & & & & & & & & \\
      \hline 
             & & & & & & & & & & \\

       PG0003+199 & 1.20 & 6.24      & 2.16      & 4.08      & 7.95      &4.56       &3.39       & 3.52     & 4.06     & 3.13     \\
             &$\pm$ 0.04 &$\pm$ 0.31 &$\pm$ 0.29 &$\pm$ 0.43 &$\pm$ 0.31 &$\pm$ 0.30 &$\pm$ 0.40 &$\pm$0.41 &$\pm$0.47 & $\pm$0.36\\
             & & & & & & & & & & \\

       Ark\,120& 1.19     & 15.09    & 4.26    & 10.82   & 18.18   & 9.02    & 9.14    & 9.41    & 10.85   & 13.27   \\
               & $\pm$0.06& $\pm$0.73&$\pm$0.59&$\pm$0.94&$\pm$0.57&$\pm$0.62&$\pm$0.84&$\pm$0.86&$\pm$0.99&$\pm$1.21\\

             & & & & & & & & & & \\
      \hline
    \end{tabular}
    \label{table2}
  \end{center}
\end{table*}

\begin{acknowledgements}
This work was supported by the Nordrhein-Westf\"alische Akademie der
Wissenschaften und der K\"unste, funded by the Federal State
Nordrhein-Westfalen and the Federal Republic of Germany, 
as well as by the CONICYT GEMINI National programme fund 32090025 for
the development of Astronomy and related Sciences. 
We thank the
director of the Calar Alto Observatory, David Barrado, for the
allocation of discretionary time and Uli Thiele and Manuel Alises for
observing the spectra. The observations on Cerro Armazones benefitted
from the care of the guardias Hector Labra, Gerard Pino, Alberto Lavin
and Francisco Arraya. Special thanks go to Roland Lemke for his
invaluable technical support, and to the anonymous referee for his 
constructive critical report. 
This research has made use of the NASA/IPAC Extragalactic Database 
(NED) which is operated by the Jet Propulsion Laboratory, California 
Institute of Technology, under contract with the National Aeronautics 
and Space Administration.
\end{acknowledgements}

%

\end{document}